\documentclass[]{spie}  

\usepackage{amsmath,amsfonts,amssymb}
\usepackage{graphicx}
\usepackage{adjustbox}
\usepackage[colorlinks=true, allcolors=blue]{hyperref}

\title{Characterization of the optical properties of the buried contact of the JWST MIRI Si:As infrared blocked impurity band detectors}

\author[a]{Ioannis Argyriou}
\author[b]{George H. Rieke}
\author[c]{Michael E. Ressler}
\author[b]{András Gáspár}
\author[a]{Bart Vandenbussche}
\affil[a]{Institute of Astronomy KU Leuven, Celestijnenlaan 200D - box 2401, 3001 Leuven, Belgium}
\affil[b]{Steward Observatory and the Department of Astronomy, The University of Arizona, 933 N Cherry Ave, Tucson, AZ, 85750, USA}
\affil[c]{Jet Propulsion Laboratory, California Institute of Technology, 4800 Oak Grove Drive, Pasadena, CA 91109, USA}

\authorinfo{Further author information: (Send correspondence to Ioannis Argyriou)\\ E-mail: ioannis.argyriou@kuleuven.be}

\begin{document}
\maketitle

\begin{abstract}
The Mid-Infrared Instrument MIRI on-board the James Webb Space Telescope uses three Si:As impurity band conduction detector arrays. MIRI medium resolution spectroscopic measurements (R$\sim$3500-1500) in the 5~$\mu m$ to 28~$\mu m$ wavelength range show a 10-30\% modulation of the spectral baseline; coherent reflections of infrared light within the Si:As detector arrays result in fringing. We quantify the shape and impact of fringes on spectra of optical sources observed with MIRI during ground testing and develop an optical model to simulate the observed modulation. We use our optical model in conjunction with the MIRI spectroscopic data to show that the properties of the buried contact inside the MIRI Si:As detector have a significant effect on the fringing behavior.
\end{abstract}


{\setlength{\parindent}{0cm}
Copyright 2020 Society of Photo‑Optical Instrumentation Engineers (SPIE).}

\section{INTRODUCTION}
\label{sec:intro}  

The Mid-Infrared Instrument MIRI on-board the James Webb Space Telescope (JWST) will perform imaging, coronagraphy, low-resolution spectroscopy, and medium-resolution spectroscopy at unprecedented sensitivity levels in the 5~$\mu m$ to 28~$\mu m$ wavelength range\cite{miri_og_paper}. MIRI will be used by astronomers to study distant galaxies, stellar populations, exoplanets, and bodies within our solar system. The instrument employs three 1024x1024 pixel Si:As IBC detector arrays, which were manufactured by Raytheon Vision Systems (RVS) \cite{love2005}.

Si:As IBC devices have extensive space flight heritage. Similar devices were used in all three instruments of the Spitzer space telescope, namely the Infrared Array Camera (IRAC), the Infrared Spectrograph (IRS), and the Multiband Imaging Photometer (MIPS). The devices were also used in the Wide-Field Infrared Survey Explorer (WISE), the Midcourse Space Experiment (MSX), the Infrared Camera (IRC) on-board Akari, and the Short Wavelength Spectrometer (SWS) on-board the Infrared Space Observatory (ISO). The high quantum efficiency of the Si:As IBC devices, in combination with the extensive wavelength range covered (5-28~$\mu m$) gives these devices a unique advantage \cite{miri_pasp_7,love2005}. Other detectors, such as the Teledyne Imaging Sensors' LWIR HgCdTe detector have a relatively higher quantum efficiency, however, they cover a shorter wavelength range (the LWIR HgCdTe detector extends out to 13~$\mu m$) \cite{Dorn2018}. It is important thus to understand the spectrophotometric response of the Si:As IBC device across the long wavelength range covered in order to maximize scientific output.

Two of the three Si:As detector arrays of MIRI are used in the medium-resolution spectrometer (MRS)\cite{miri_pasp_6}. The detectors sample the 5~$\mu m$ to 28~$\mu m$ wavelength range. The MRS spectral resolution varies from $\lambda/\Delta\lambda\sim$ 3500 at 5~$\mu m$ and $\lambda/\Delta\lambda\sim$ 1500 at 28~$\mu m$ (Labiano, Argyriou, et al., 2021, in prep). As is common for infrared spectrometers, constructive and destructive interference in different layers of the MRS detector arrays modulate the detected signal as a function of wavelength; the detector-constituting layers behave as efficient Fabry-P\'erot etalons. The resulting "fringing" in the MRS spectra varies in amplitude between 10-30\% of the spectral baseline\cite{argyriou2020}. Similar fringing was observed in ISO SWS data, in Spitzer IRS data, as well as near-infrared data taken with the Space Telescope Imaging Spectrograph (STIS) on-board the Hubble Space Telescope (HST) \cite{Kester2003,lahuis2003,stis_fringing_malumuth}. If not corrected, fringing can severely hamper the scientific interpretation of MIRI observations.

As we will show in this paper, fringing in the MRS detectors is measured as an apparent beating of two fundamental frequencies. The first frequency is proportional to the geometric thickness between the detector back side entrance surface and the pixel metalization. The second frequency is proportional to the geometric thickness between the buried contact, characteristic of back-illuminated Si:As IBC devices, and the pixel metalization. The first frequency has been studied extensively in [\citenum{argyriou2020}]. Similar to MRS data, [\citenum{woods2011}] identified a second fringe in their study of Si:As IBC devices. In this paper we characterize the second fringe in the MIRI Si:As devices using the resolved fringes in MRS spectra. More importantly, we argue that the physical cause of the second fringe is the high doping concentration of the buried contact. 

\section{MIRI DETECTOR ARCHITECTURE AND MRS SPECTRAL FRINGING}
\begin{figure}[h]
\begin{center}
\begin{tabular}{c}
\includegraphics[width=0.32\textwidth]{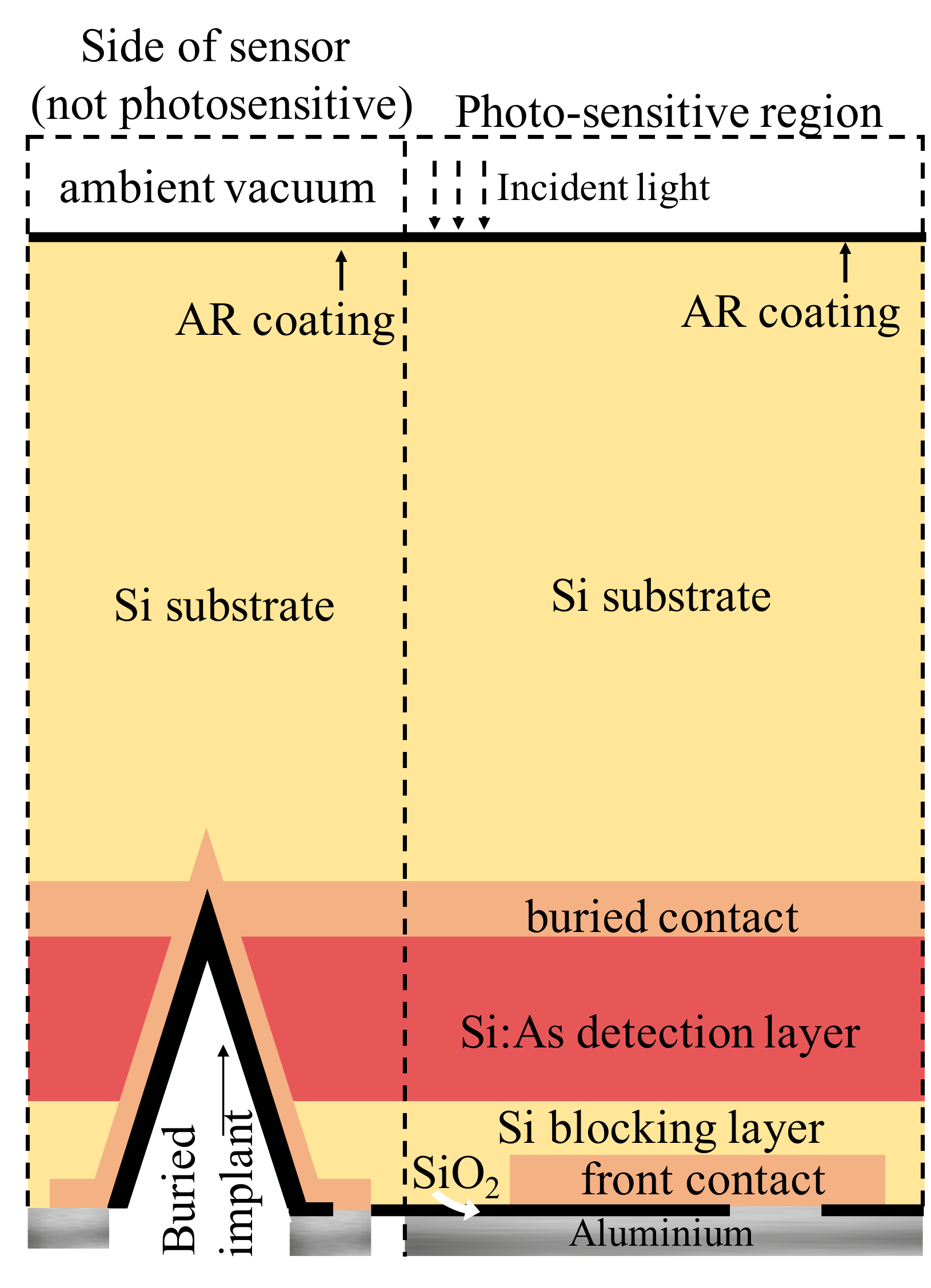}
\end{tabular}
\end{center}
\caption[example]{Representation of MIRI detector architecture based on [\citenum{petroff_stapelbroek_patent,love2005,miri_pasp_7,gaspar2020quantum}]. The dimensions are not to scale.}
\label{fig:detector_layout}
\end{figure}

The Si:As IBC detectors are grown on a silicon substrate. We show a representation of the architecture of the MIRI detector arrays in Fig.~\ref{fig:detector_layout}, which is based on [\citenum{petroff_stapelbroek_patent,love2005,miri_pasp_7,gaspar2020quantum}]. Photons pass through the anti-reflection coating on the detector back side, into the substrate, and then through the buried contact, into the infrared-active layer (detection layer). The infrared-active layer is doped with arsenic to absorb the incoming photons, elevating photo-excited electrons from the impurity band into the conduction band\cite{miri_pasp_7}. Assuming a low level of minority acceptor impurities in this layer, an electric field can be maintained across it that causes the photoelectrons to migrate to the front of the detector. A thin, high purity layer (blocking layer) is grown over the front of the infrared-active layer. When operated at sufficiently low temperatures, thermally-generated free charge carriers cannot penetrate the blocking layer because it lacks an impurity band and the carriers have insufficient energy to be lifted into the conduction band. The photo-generated free charge carriers in the conduction band can, however, traverse the blocking layer, to be collected at the front contact. The electric field that drives this process is maintained across the infrared-active layer between the front contact, on the readout side, and a buried contact, connected via a V-etch buried implant on one side of the detector.

\newpage

The MIRI MRS has a short-wavelength arm (SW) and a long-wavelength arm (LW)\cite{miri_pasp_6}. The SW and LW arms use one Si:As IBC device each, which detect photons in the 5-11.8~$\mu m$ range and in the 11.8-28~$\mu m$ range respectively. We provide the optical and geometric properties of the MRS SW and LW detector layers in Table \ref{tab:miri_sw_detector_optical_layers} and Table \ref{tab:miri_lw_detector_optical_layers}\footnote{A semi-infinite thickness is attributed to the ambient medium and the pixel metalization.}. These properties define the optical medium through which the photons propagate.

\begin{table}[h]
\caption{ Optical and geometric properties of MIRI MRS SW detector layers\cite{love2005,argyriou2020}.}
\begin{adjustbox}{scale=0.9}
\centering
\begin{tabular}{cccl}
\hline\hline
Layer    & Material & Geometric thickness & Comments \\
\hline
Ambient   & Vaccuum  & inf. &  \\
Anti-reflection coating   & Zinc Sulphide (ZnS)  & 0.72~$\mu m$ & Minimum reflectance at 6~$\mu m$. \\
Inactive layer & Silicon (Si)  &   460.5$\pm$6.6~$\mu m$ & Pure silicon (zero absorption). \\
Buried contact (electrode)  & Antimony-doped silicon & uncertain & Material is assumed based on [\citenum{woods2011}]. \\
Active layer  & Arsenic-doped silicon   &  30~$\mu m$ & - Thickness variation much less than $1\mu m$.\\
 &    &   & - Arsenic concentration N=$5 \times 10^{17} cm^{-3}$.\\
  &    &   & - Contingency array (identical to IRAC).\\
Blocking layer & Silicon (Si)  &  4~$\mu m$ & Pure silicon (zero absorption). \\
Pixel metalization & Aluminium (Al)  &  inf. &  \\
\hline
\label{tab:miri_sw_detector_optical_layers}
\end{tabular}
\end{adjustbox}
\end{table}

\begin{table}[h]
\caption{Optical and geometric properties of MIRI MRS LW detector layers\cite{love2005,argyriou2020}.}
\begin{adjustbox}{scale=0.9}
\centering
\begin{tabular}{cccl}
\hline\hline
Layer    & Material & Geometric thickness & Comments \\
\hline
Ambient   & Vaccuum  & inf. &  \\
Anti-reflection coating   & Zinc Sulphide (ZnS)  & 2.08~$\mu m$ & Minimum reflectance at 16~$\mu m$. \\
Inactive layer & Silicon (Si)  &   416.8$\pm$5.2~$\mu m$ & Pure silicon (zero absorption). \\
Buried contact (electrode)  & Antimony-doped silicon & uncertain & Material is assumed based on [\citenum{woods2011}]. \\
Active layer  & Arsenic-doped silicon   &  35~$\mu m$ & - Thickness variation much less than $1\mu m$.\\
 &    &   & - Arsenic concentration N=$7 \times 10^{17} cm^{-3}$.\\
 &    &   & - Baseline array.\\
Blocking layer & Silicon (Si)  &  4~$\mu m$ & Pure silicon (zero absorption). \\
Pixel metalization & Aluminium (Al)  &  inf. &  \\
\hline
\label{tab:miri_lw_detector_optical_layers}
\end{tabular}
\end{adjustbox}
\end{table}

Since the infrared-active layer is doped with arsenic, an n-type impurity, a similar dopant will have been used for the buried contact, i.e. either arsenic (As), antimony (Sb), or phosphorus (P). Antimony and phosphorus are two commonly used n-type dopants in semiconductors\cite{semiconductor_tech}. In fact, Si:As IBC devices manufactured by the Rockwell Science Center (USA) used antimony for the buried contact and arsenic for the front contact\cite{woods2011}. We assume that that is the case for the MIRI detectors as well.

\begin{figure}[t]
\begin{center}
\begin{tabular}{c}
\includegraphics[width=0.97\textwidth]{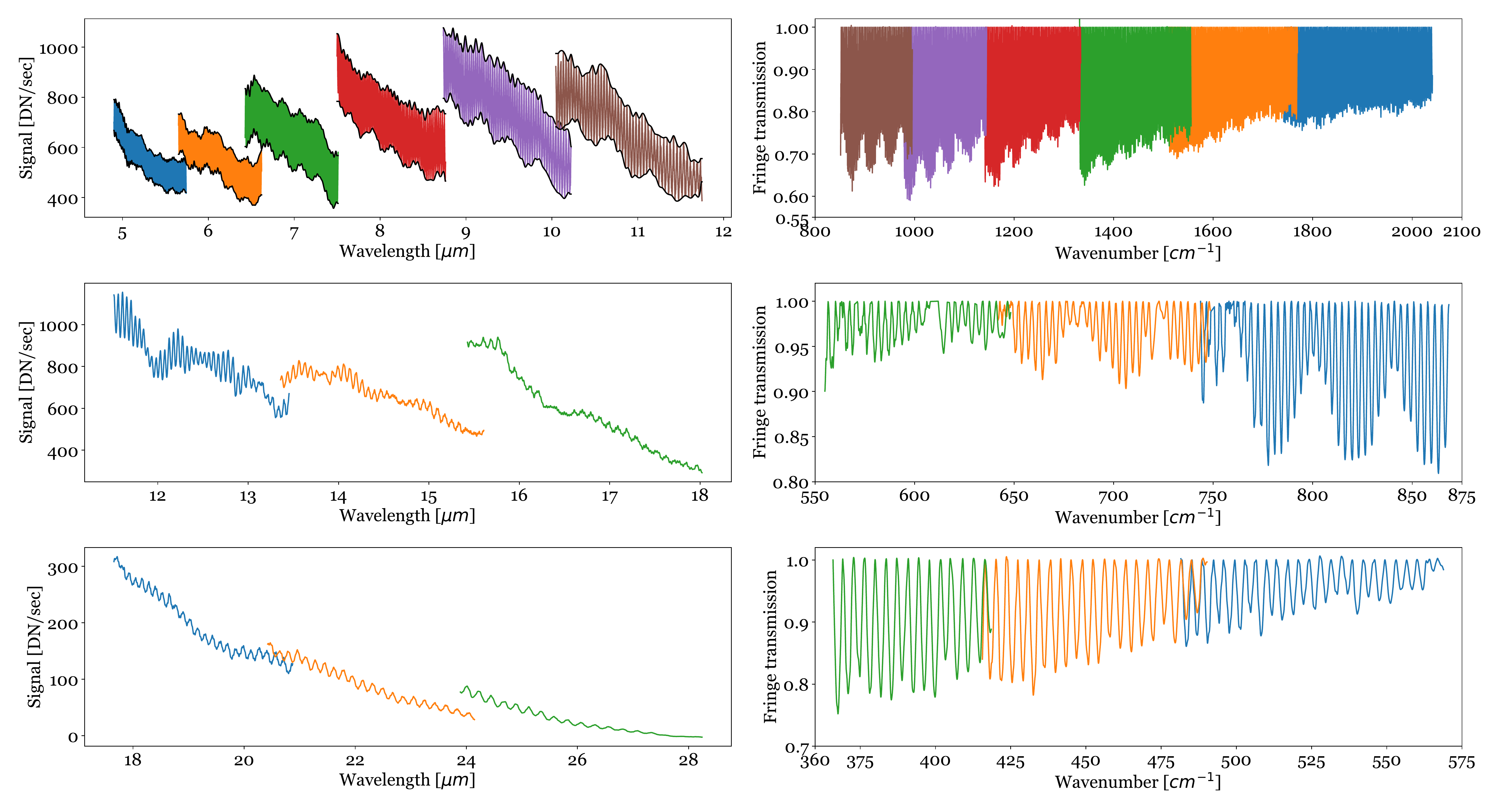}
\end{tabular}
\end{center}
\caption[example]{Left panel: MRS extracted spectra, as a function of wavelength, from observations of the spectrometer calibration source. The top plot shows the SW data, and the middle and bottom plots show the two halves of the LW wavelength range. Right panel: Fringe transmission as a function of wavenumber. The transmission is determined by normalizing the spectra in the left panel to a fringe free continuum (spline connecting the fringe peaks). Transmission discontinuities between nearby spectral bands in the SW data are due to each band having a different spectral resolution.}
\label{fig:low_frequency_modulation}
\end{figure}

The MRS acquires spectra in 12 spectral bands, 6 with the SW detector and 6 with the LW detector. In the left panel of Fig.~\ref{fig:low_frequency_modulation} we show MRS measurements taken with the spectrometer calibration source in all 12 MRS spectral bands\footnote{The MRS calibration source is a spatially flat source which emits as a grey body at a temperature of 800K\cite{miri_pasp_6}}. The x-axis shows the wavelength coordinate and the y-axis gives the signal units in digital numbers (DN) per second. We note that no spectrophotometric correction has been applied to the data. Discontinuities in the signal of neighboring spectral bands are due to jumps in the spatial and spectral resolution of each band, which also causes the systematic drop in signal within each band. 

Focusing on the top plot of the left panel of Fig.~\ref{fig:low_frequency_modulation}, we see a prominent fringe envelope in all 6 of the SW bands showcased. This fringing is caused by the coherent reflections between the detector back side surface coating and the pixel metalization. The MRS SW anti-reflection coating is optimized such that reflections between the ambient medium and the silicon substrate are least at 6~$\mu m$, however, the reflectance of the coating never drops below 5\% (a constant reflectance of 98\% can be assumed at the pixel aluminum metalization). This results in a prominent high-frequency sinusoidal signature at all wavelengths below 12~$\mu m$. Superimposed on the high-frequency fringes is a low frequency modulation, a second fringe. The modulation in question can be discerned most clearly as a function of wavelength when looking at the lower part of the envelope of the 6 SW spectra. The picture becomes slightly less clear in the middle plot (LW bands), where different detector parameters between the SW and LW bands give a discontinuous image with respect to the top plot. There is, however, a clear beating pattern in the signal. This beating pattern is no longer resolved in the bottom plot, where we see mainly the primary fringe frequency, but no statistically significant sign of the secondary frequency.

A common way to study fringes is to plot them in wavenumber ($\sigma=1/\lambda$) space. This is shown in the right panel of  Fig.~\ref{fig:low_frequency_modulation}, where we normalize the signal shown in the left panel to a spline connecting the fringe peaks, which we define as the "fringe-free" spectral continuum. Going from high to low wavenumbers, we see that the fringe beating between the high and low frequency fringes becomes more prominent from 2050~$cm^{-1}$ ($\sim$4.88~$\mu m$) down to 610~$cm^{-1}$ ($\sim$16.4~$\mu m$). At even lower wavenumbers, below 575~$cm^{-1}$, there is a lack of a clearly resolved beating pattern.

Assuming that the layers in the MRS detectors are plane parallel, the distance between interference peaks in a fringe transmission spectrum can be related to the optical thickness of the Fabry-P\'erot etalon as $D=1/\Delta\sigma$. The optical thickness is related to the geometric thickness of the etalon $t_{et} = (D/2)/n$, where $n$ is the refractive index of the medium between the etalon reflective boundaries. We use the optical properties of crystalline silicon for the Si:As IBC device\cite{bradley2006}. The relation for $D$ and $t_{et}$ can be used to probe the optical parameters of the etalon producing the MRS fringes. A more advanced method to determining the optical thickness $D$ is to fit the fringe transmission with a Fabry-P\'erot transmittance function, given by Eq.~\ref{eq:transmittance function} \cite{lipson1969,argyriou2020}. In this equation $R$ stands for the equivalent reflectance of the etalon, assuming that both etalon boundaries have the same reflectance. We use a scanning window with a fixed width (two fringe periods) and fit the fringes inside the specified window as we scan the window across the spectrum.

\begin{equation}\label{eq:transmittance function}
    T_{et}(\sigma) = \left(1 + \frac{4R}{(1-R)^2} \sin^2\left(2\pi D\sigma\right) \right)^{-1}
\end{equation}

\begin{figure}[t]
\begin{center}
\begin{tabular}{c}
\includegraphics[width=0.6\textwidth]{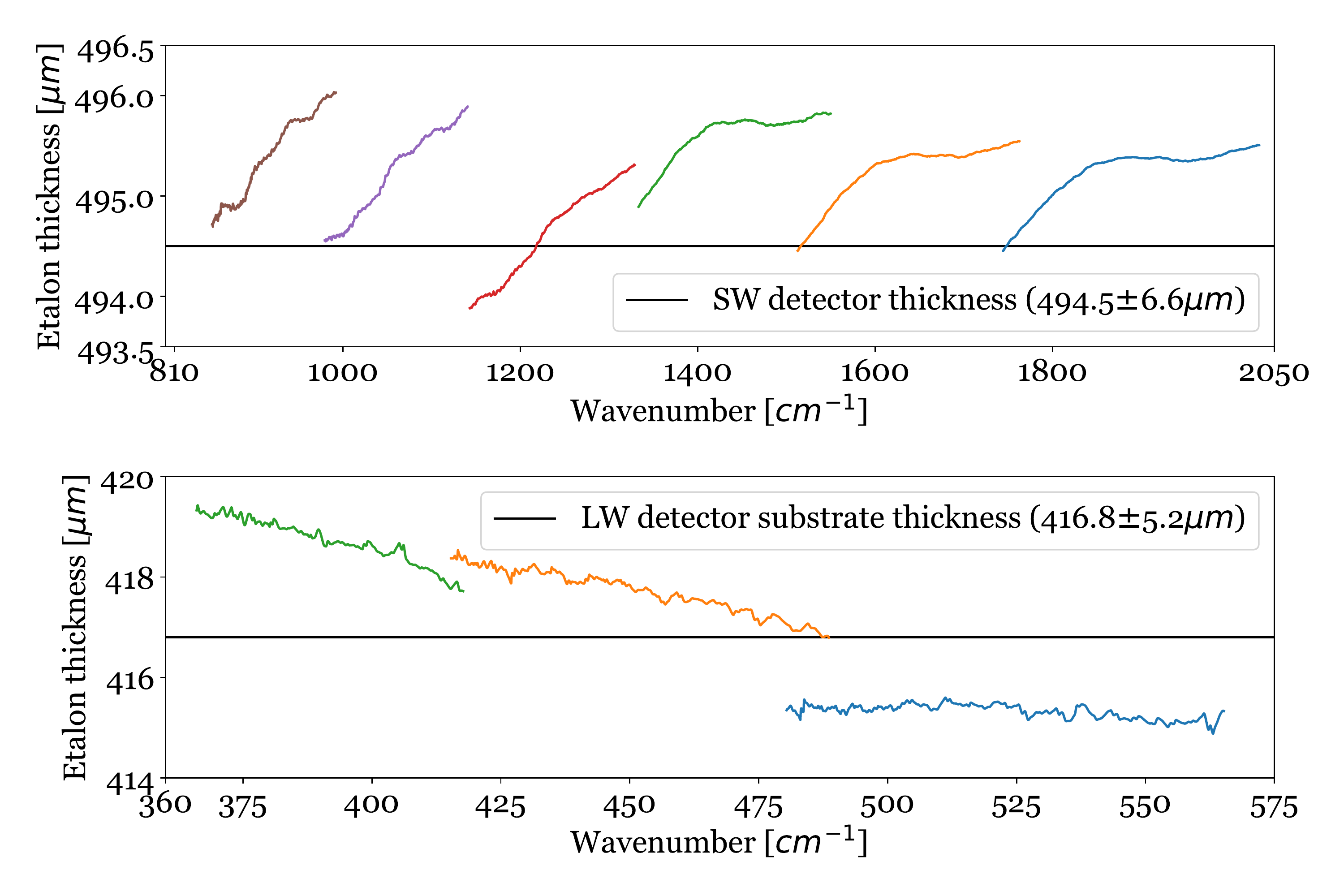}
\end{tabular}
\end{center}
\caption[example]{Etalon thickness derived from fitting 9 fringe transmission profiles (6 in SW data, 3 in LW data) as a function of wavenumber. Each wavenumber corresponds to a different pixel location on the MRS detectors, which allows us to probe detector thickness variations.}
\label{fig:etalon_thickness}
\end{figure}

In Fig.~\ref{fig:etalon_thickness} we show the resulting etalon thickness $t_{et}$, estimated from the fringe transmission in the top and bottom right plot of Fig.~\ref{fig:low_frequency_modulation}. From fitting the high-frequency fringe in the SW data, we find that the etalon thickness matches the sum of the SW detector back side surface (inactive layer), infrared-active layer, and blocking layer. We assume that the thickness of the buried contact layer is much smaller compared to the other layers. Consequently, this means that the fringe is produced between the back side surface coating and the pixel metalization. Using the low-frequency fringe from SW data, a rough back-of-the-envelope calculation yields a thickness of 36.9$\pm$3~$\mu m$ (uncertainty based on the standard deviation of the limited measurements). We find that this is in agreement with the infrared-active layer plus blocking layer thickness (34~$\mu m$). In the bottom plot of Fig.~\ref{fig:etalon_thickness}, we find that the frequency of the fringing in the bottom plot of Fig.~\ref{fig:low_frequency_modulation} matches that of the back side surface (inactive layer) of the LW detector; the fringing occurs between the back side surface coating and the infrared-active layer.

\section{MIRI DETECTOR OPTICAL MODEL}

Each layer inside the MIRI Si:As IBC devices has optical properties defined by the materials listed in Table~\ref{tab:miri_sw_detector_optical_layers} and \ref{tab:miri_lw_detector_optical_layers}. We use refractive index values for zinc-sulphide, silicon, and aluminium, based on the available literature\cite{Amotchkina20,bradley2006,ordal_1988}. Absorption inside the infrared-active layer is parametrized using the absorption coefficient curve of Si:As derived by [\citenum{woods2011}]\footnote{We fit the 5-20~$\mu m$ data with a third-order polynomial and extrapolate the trend to 28~$\mu m$. The result is in agreement with the values of [\citenum{Geist89,gaspar2020quantum}].}.

We model the transmitted, reflected, and absorbed intensity of the MIRI detectors by using the refractive information of each material, and solving the Fresnel equations analytically for a multilayer planar stack. We do this using the Transfer-Matrix Method (TMM)\cite{born1970,byrnes_tmm}. What we measure in MRS data is the fraction of the light absorbed in the infrared-active layer at each wavelength. We focus thus on the absorbed intensity prediction of the TMM simulation.

\begin{figure}[t]
\begin{center}
\begin{tabular}{c}
\includegraphics[width=0.7\textwidth]{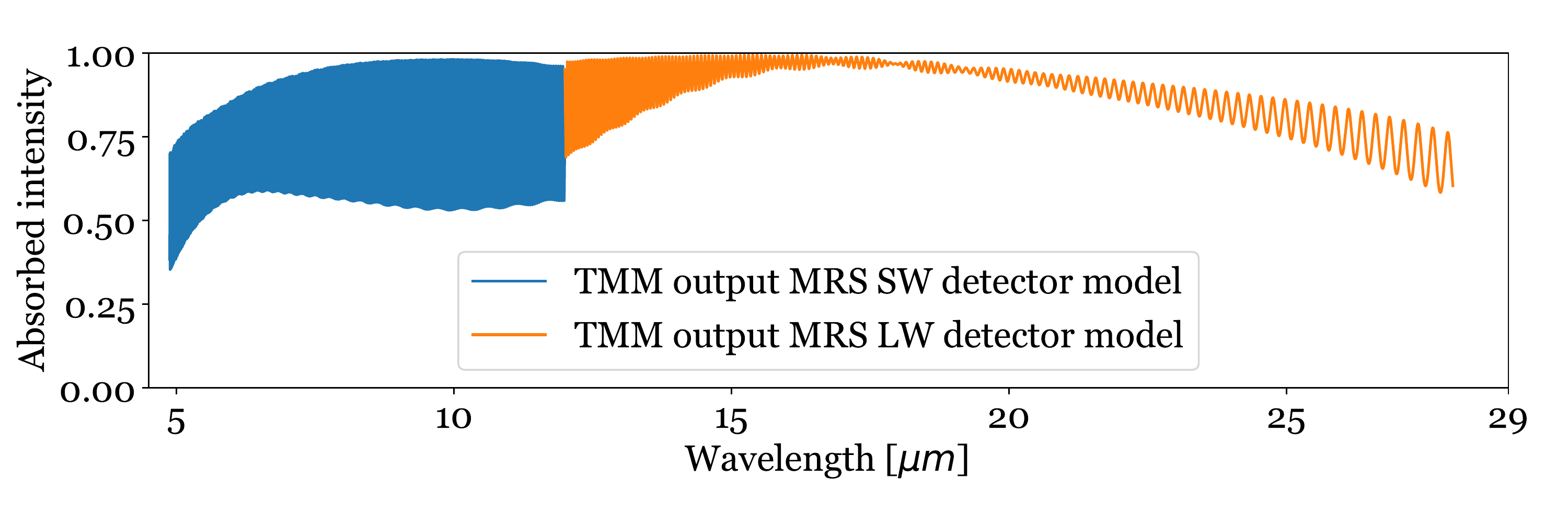}
\end{tabular}
\end{center}
\caption[example]{Absorbed intensity simulated using the Transfer-Matrix Method (TMM) for the MRS SW and LW detector optical models. The buried contact is not accounted for in the optical models.}
\label{fig:tmm_simulation_basic_model}
\end{figure}

In Fig.~\ref{fig:tmm_simulation_basic_model} we show the result of a pair of high-resolution TMM simulations based on the MRS SW and LW detector optical models. The traditional assumption has been that the buried contact is transparent and does not affect the optical performance, so for these simulations no buried contact was included in the models. The simulations show a beating of two frequencies, similar to the MRS data shown in Fig.~\ref{fig:low_frequency_modulation}. However, we find a significant discrepancy between the TMM simulated fringes and the fringes in the MRS data, namely, the amplitude of the secondary frequency modulation in the TMM output is much smaller compared to the MRS data. This is noticeable in the SW regime, as well as in the LW regime. In the latter case, there is a significant difference in the 12-18.2~$\mu m$ wavelength range (830-550~$cm^{-1}$). The beating in the LW data is very prominent already at 12~$\mu m$ (830~$cm^{-1}$), which is not the case for the simulated absorbed intensity.

All material properties in the MRS optical models are well constrained, except for the material properties of the buried contact layer. We can probe the material properties of the buried contact layer by minimizing the difference between the fringes in the TMM-simulated absorbed intensity and the MRS fringes. This is done as follows:
\begin{enumerate}
    \item We assume a thickness of the buried contact layer of 0.2~$\mu m$ (typical value for an ion implanted layer).
    \item We define the buried contact layer refractive index as $n_{bc} = n_{Si}+\Delta n_{bc}$, where $n_{Si}$ is the refractive index of crystalline silicon\cite{bradley2006}. $\Delta n_{bc}$ is defined as a free parameter. For simplicity, the value of $\Delta n_{bc}$ is kept constant in each MRS spectral band.
    \item We assume that the absorption coefficient of the buried contact layer is the same as that of the infrared-active layer. Given the small thickness of the layer (0.2~$\mu m$ versus 30-35~$\mu m$ for the active layer), a difference in the absorption coefficient between the buried contact and the infrared-active layer is a second order effect.
    \item We keep all other refractive and geometric model properties fixed, defined as per Table~\ref{tab:miri_sw_detector_optical_layers} and Table~\ref{tab:miri_lw_detector_optical_layers}.
    \item We perform a TMM simulation at each evaluation of the $\Delta n_{bc}$ parameter.
    \item We convolve the simulation results by the MRS spectral resolution.
    \item We bring the optical model and the MRS fringes on the same comparative baseline. This is done by determining a running mean through both datasets and dividing by this running mean.
    \item We compute the value of $\Delta n_{bc}$ that minimizes the difference between the TMM-simulated fringes and the MRS fringes in each of the MRS spectral bands.
\end{enumerate}

\section{RESULTS AND DISCUSSION}

\begin{figure}[h]
\begin{center}
\begin{tabular}{c}
\includegraphics[width=0.75\textwidth]{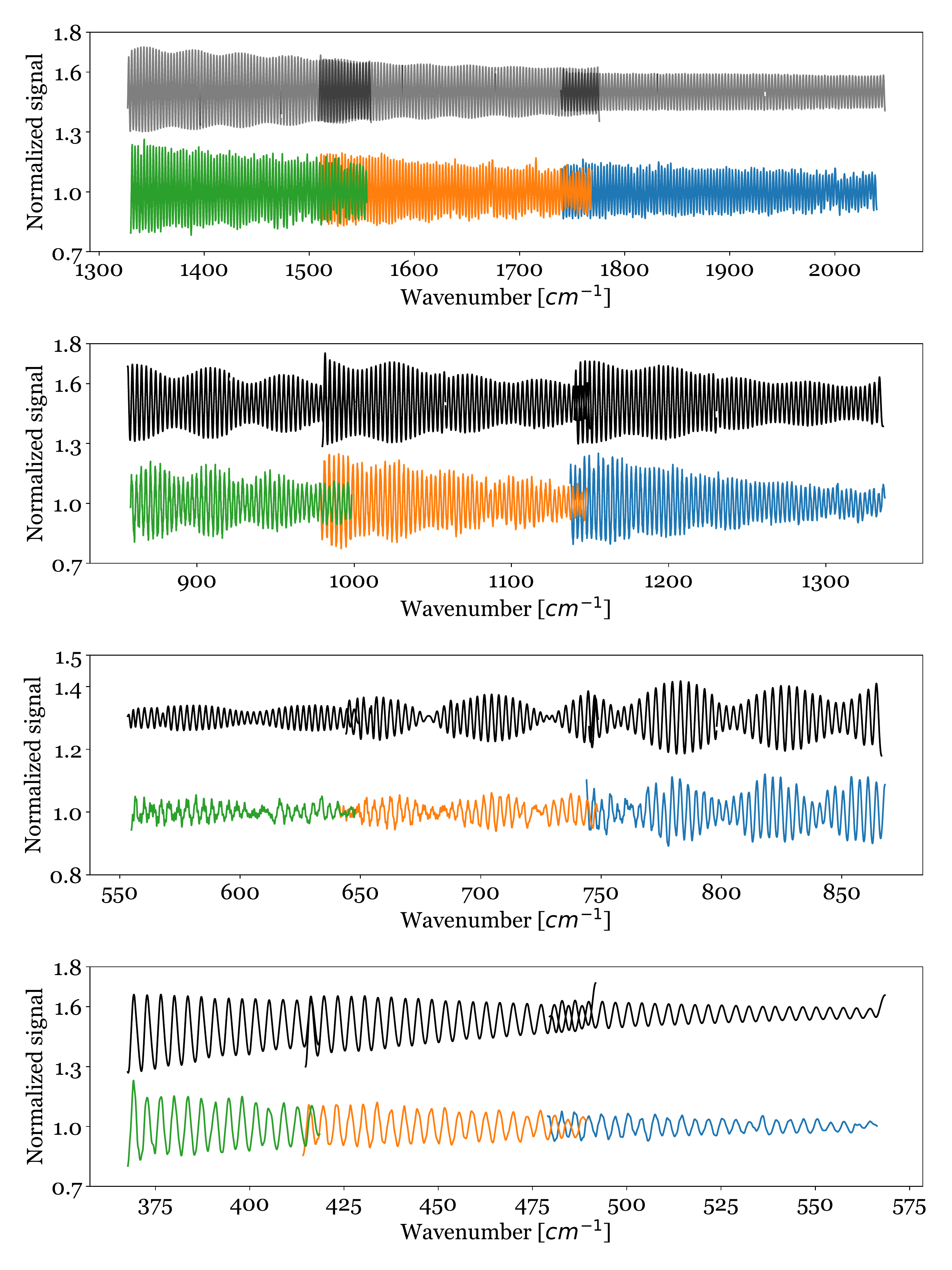}
\end{tabular}
\end{center}
\caption[example]{Minimization analysis result. The black data show the TMM models based on an optimized value for $\Delta n_{bc}$.}
\label{fig:optimized_tmm_vs_data}
\end{figure}

\begin{figure}[t]
\begin{center}
\begin{tabular}{c}
\includegraphics[width=0.4\textwidth]{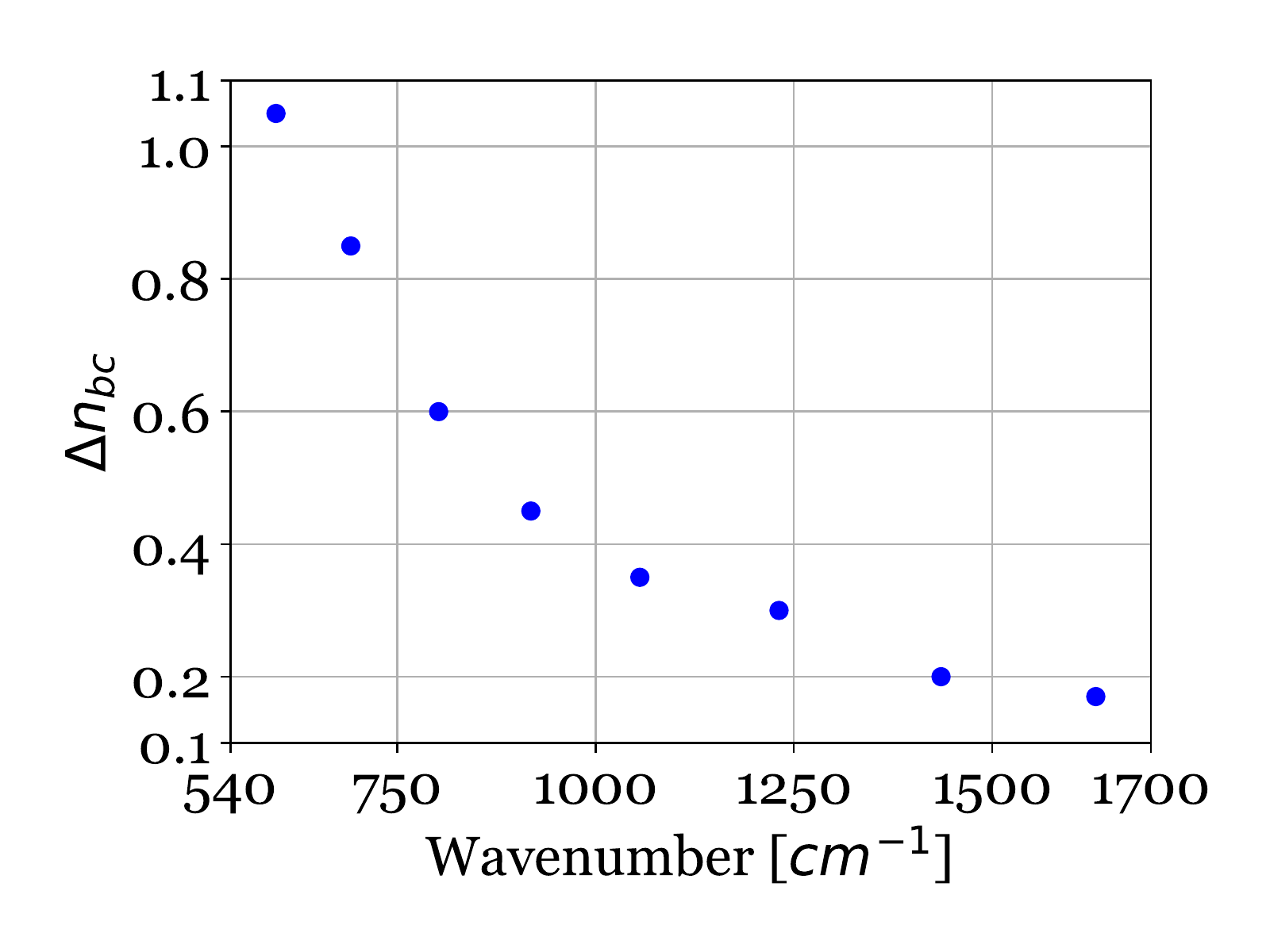}
\end{tabular}
\end{center}
\caption[example]{Optimal values for $\Delta n_{bc}$ determined from minimizing the difference between the MRS TMM simulation fringes and fringes in MRS data.}
\label{fig:refractive_index_offset}
\end{figure}

We show the results of our analysis in Fig.~\ref{fig:optimized_tmm_vs_data} and Fig.~\ref{fig:refractive_index_offset}. In Fig.~\ref{fig:optimized_tmm_vs_data} the black data show the TMM simulation based on the optimized value of $\Delta n_{bc}$. The colored data show the MRS fringes. The optimized values for $\Delta n_{bc}$ as a function of wavelength are shown in Fig.~\ref{fig:refractive_index_offset}. We report a significant change in the refractive index of the buried contact as a function of wavenumber.


The index of refraction of a solid such as doped silicon can be expressed analytically using the classical dispersion theory of solids\cite{moss1973}. [\citenum{Zhou1994}] gives a theoretical formulation for this. We note that we found an erratum (sign error) in equation 12 of [\citenum{Zhou1994}]. The correct expression for the index of refraction is given in Eq.~\ref{eq:zhou_correct_equation_12}. In this equation, $\epsilon^{*}$ is the permittivity of silicon, $\omega_{p}$ is the plasma frequency (in units of Hz), $\omega$ is the wave angular frequency ($\omega = 2\pi c /\lambda $, in units of Hz, where $c$ is the speed of light), and $g$ is the damping coefficient. The plasma frequency $\omega_{p}$ is proportional to the doping concentration N, and the damping coefficient $g$ is inversely proportional to the carrier mobility $\mu$ (all the relevant equations and parameters can be found in [\citenum{Zhou1994}]). The carrier mobility $\mu$ depends on the semiconductor temperature and on the doping concentration \cite{ibach2009,vanZeghbroeck2011}. The MIRI detectors are operated at cryogenic temperatures ($\sim$7~K). In this regime the carrier mobility is approximately constant as a function of temperature, as such it is mainly dependent on the doping concentration. For the doping we consider the carrier mobility of degenerate electrons in antimony-doped silicon\cite{Norton1973,Canali1975,rieke2003}.

\begin{equation}\label{eq:zhou_correct_equation_12}
\begin{aligned}
n^{2}=& \frac{\epsilon^{*}}{2}\left\{\left(1-\frac{\omega_{p}^{2}}{\omega^{2}+g^{2}}\right)+\left[\left(1-\frac{\omega_{p}^{2}}{\omega^{2}+g^{2}}\right)^{2}+\left(\frac{g \omega_{p}^{2}}{\omega\left(\omega^{2}+g^{2}\right)}\right)^{2}\right]^{1 / 2}\right\}
\end{aligned}
\end{equation}

\begin{figure}[h]
\begin{center}
\begin{tabular}{c}
\includegraphics[width=0.5\textwidth]{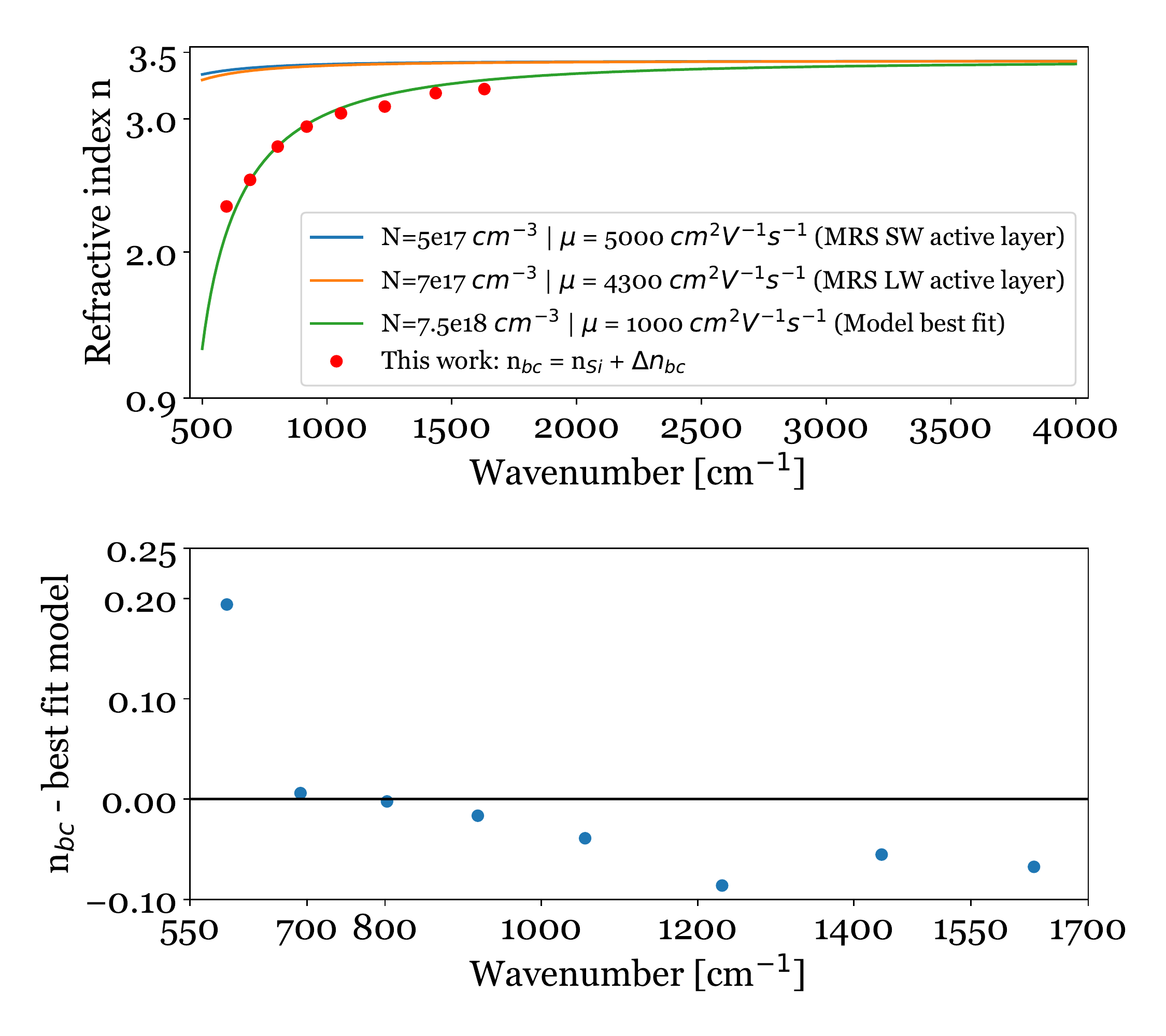}
\end{tabular}
\end{center}
\caption[example]{Top: Index of refraction of the buried contact $n_{bc}$ computed in this work, the best fit to the determined $n_{bc}$ values based on Eq.~\ref{eq:zhou_correct_equation_12}, and the refractive index based on Eq.~\ref{eq:zhou_correct_equation_12} for the MRS SW and LW detectors, computed using the doping concentration of the respective infrared-active layers. Bottom: Residals between $n_{bc}$ and the best fit model.}
\label{fig:resulting_index_of_refraction}
\end{figure}

In the top plot of Fig.~\ref{fig:resulting_index_of_refraction} we plot the index of refraction of the buried contact $n_{bc}$, computed in this work, the best fit to the determined $n_{bc}$ values based on Eq.~\ref{eq:zhou_correct_equation_12}, and the refractive index for the MRS SW and LW detectors based on Eq.~\ref{eq:zhou_correct_equation_12}. For the latter we use the doping concentration of the respective infrared-active layers given in Table~\ref{tab:miri_sw_detector_optical_layers} and \ref{tab:miri_lw_detector_optical_layers}. The plotted curves are consistent with [\citenum{Hava2007}]. The index of refraction of the MRS SW and LW infrared-active layer is approximately constant down to 1000~$cm^{-1}$, below which a small deviation is predicted. For the buried contact we find a best fit model with doping concentration N=$7.5\times 10^{18}cm^{-3}$ and carrier mobility $\mu = 1000cm^2/(V\cdot s)$, which is consistent with [\citenum{chow1966}].

In the bottom plot of Fig.~\ref{fig:resulting_index_of_refraction} we show the residuals of subtracting the best fit model from the $n_{bc}$ values. We identify three contributions to the residuals:
\begin{itemize}
    \item Different doping in the MRS SW and LW detectors: Although we perform a single fit to all the $n_{bc}$ values, there may be a small difference in the doping of the MRS SW and LW detectors.
    \item Only a single $\Delta n_{bc}$ value was determined per MRS spectral band. The variation in index of refraction within a band is expected to be gradual instead of a step function.
    \item Uncertainties in the modeling of the MRS high-frequency fringing. A sub-optimal correction of the high-frequency fringing results in residuals in the low-frequency fringe profile, which impact the estimation of $\Delta n_{bc}$.
\end{itemize}


\section{CONCLUSIONS AND FUTURE WORK}
We have shown that JWST MIRI MRS spectra contain two dominant frequencies in their spectra. By fitting the resolved fringes with a Fabry-P\'erot transmittance function, we find that the high frequency fringing is produced between the detector back side surface coating and the pixel metalization. The low frequency fringing is the result of a beating between the high-frequency fringing and fringing produced between the detector back side surface and the infrared-active layer. By using the Transfer-Matrix Method and an optical model for the MRS SW and LW detectors, we show that the optical behavior of the buried contact is consistent with it having a doping concentration of N=$7.5\times 10^{18}cm^{-3}$. This doping concentration results in a significant change in the index of refraction at the buried contact layer. We argue that this accounts for the behavior of the prominent low-frequency fringing in MIRI MRS data. 

In future work, the impact of the buried contact layer thickness on the MRS fringes will be investigated. A larger thickness results in more absorption inside the layer, which affects the fringing. Due to the relatively small size of the buried contact layer, compared to the infrared-active layer, absorption in the buried contact is a second order effect. Nevertheless, a complete model of the detector requires fitting both refraction and absorption effects simultaneously.

\acknowledgments 
 
Ioannis Argyriou and Bart Vandenbussche thank the European Space Agency (ESA) and the Belgian Federal Science Policy Office (BELSPO) for their support in the framework of the PRODEX Programme.

The work presented is the effort of the entire MIRI team and the enthusiasm within the MIRI partnership is a significant factor in its success. MIRI draws on the scientific and technical expertise of the following organisations: Ames Research Center, USA; Airbus Defence and Space, UK; CEA-Irfu, Saclay, France; Centre Spatial de Liége, Belgium; Consejo Superior de Investigaciones Científicas, Spain; Carl Zeiss Optronics, Germany; Chalmers University of Technology, Sweden; Danish Space Research Institute, Denmark; Dublin Institute for Advanced Studies, Ireland; European Space Agency, Netherlands; ETCA, Belgium; ETH Zurich, Switzerland; Goddard Space Flight Center, USA; Institute d'Astrophysique Spatiale, France; Instituto Nacional de Técnica Aeroespacial, Spain; Institute for Astronomy, Edinburgh, UK; Jet Propulsion Laboratory, USA; Laboratoire d'Astrophysique de Marseille (LAM), France; Leiden University, Netherlands; Lockheed Advanced Technology Center (USA); NOVA Opt-IR group at Dwingeloo, Netherlands; Northrop Grumman, USA; Max-Planck Institut für Astronomie (MPIA), Heidelberg, Germany; Laboratoire d’Etudes Spatiales et d'Instrumentation en Astrophysique (LESIA), France; Paul Scherrer Institut, Switzerland; Raytheon Vision Systems, USA; RUAG Aerospace, Switzerland; Rutherford Appleton Laboratory (RAL Space), UK; Space Telescope Science Institute, USA; Toegepast- Natuurwetenschappelijk Onderzoek (TNO-TPD), Netherlands; UK Astronomy Technology Centre, UK; University College London, UK; University of Amsterdam, Netherlands; University of Arizona, USA; University of Bern, Switzerland; University of Cardiff, UK; University of Cologne, Germany; University of Ghent; University of Groningen, Netherlands; University of Leicester, UK; University of Leuven, Belgium; University of Stockholm, Sweden; Utah State University, USA. A portion of this work was carried out at the Jet Propulsion Laboratory, California Institute of Technology, under a contract with the National Aeronautics and Space Administration.

We would like to thank the following National and International Funding Agencies for their support of the MIRI development: NASA; ESA; Belgian Science Policy Office; Centre Nationale D'Etudes Spatiales (CNES); Danish National Space Centre; Deutsches Zentrum fur Luft-und Raumfahrt (DLR); Enterprise Ireland; Ministerio De Economiá y Competividad; Netherlands Research School for Astronomy (NOVA); Netherlands Organisation for Scientific Research (NWO); Science and Technology Facilities Council; Swiss Space Office; Swedish National Space Board; UK Space Agency.

We take this opportunity to thank the ESA JWST Project team and the NASA Goddard ISIM team for their capable technical support in the development of MIRI, its delivery and successful integration.

\bibliography{report} 
\bibliographystyle{spiebib} 

\end{document}